\documentclass[%
 reprint,
showpacs,preprintnumbers,
 amsmath,amssymb,
 aps,
pra,
]{revtex4-1}

\usepackage{gensymb}
\usepackage[colorlinks=false, pdfborder={0 0 0}]{hyperref}
\usepackage[caption=false]{subfig}
\usepackage{amssymb}
\usepackage{amsmath}
\usepackage{commath}
\usepackage{graphicx,bm}
\usepackage{verbatim}
\usepackage{booktabs}

\usepackage{color,soul}

\usepackage{dcolumn}
\usepackage{bm}

\graphicspath{ {Figures/}}

\begin{document}

\preprint{APS/}

\title{Triggering filamentation using turbulence}

\author{D. Eeltink}
\author{N. Berti}
\author{N. Marchiando}
\author{S. Hermelin}
\author{J. Gateau}
\author{M. Brunetti}
\author{J.P. Wolf}
\author{J. Kasparian} 
\affiliation{Universit\'e de Gen\`eve, GAP, Chemin de Pinchat 22, 1211 Geneva 4, Switzerland}
\date{\today}%

\begin{abstract}
We study the triggering of single filaments due to turbulence in the beam path for a laser of power below the filamenting threshold. Turbulence can act as a switch between the beam not filamenting and producing single filaments. This 'positive' effect of turbulence on the filament probability, combined with our observation of off-axis filaments suggests the underlying mechanism is modulation instability caused by transverse perturbations. We hereby experimentally explore the interaction of modulation instability and turbulence, commonly associated with multiple-filaments, in the single-filament regime.
\end{abstract}


\maketitle

\section{\label{sec:Introduction}Introduction}

Filamentation denotes the self-guided propagation regime of high power, ultra-short laser pulses \cite{Couairon2007,Berge2007,Chin2010}. When the power of a laser pulse overcomes a critical nonlinear threshold $P_{NL}$ the response of the medium causes the beam to focus itself instead of simply diffracting: the Kerr-effect. The self-guidance of the pulse relies on a balancing act between the focusing Kerr-effect, and a defocusing caused by ionization and higher order Kerr-effects \cite{Loriot2011}, and can last for tens of meters, creating high intensity, small diameter ($\sim$ 100 $\mu$m) 'bullets of light' \cite{Couairon2007}. Filaments can be initiated from hundreds of meters away \cite{Rodriguez2004} and as such are promising candidates for innovative applications in several atmospheric fields such as remote sensing \cite{Kasparian2003} or laser-induced condensation \cite{Henin2011,Mongin2015} 

An important factor to take into account in these applications is the interaction of the laser pulse with a perturbed atmosphere, both before and during filamentation. Especially when the application relies on low power and single-filaments, this interaction is crucial as it can mean the difference between absence or presence of filaments. 

Turbulence causes a fluctuation of air density, and thus a fluctuation of the refractive index of the air. The interaction with the laser pulse is two-fold, affecting the beam wander and its transverse coherence length. On the one hand, it is characterized by the lensing effect of the refractive index curvature, characterized by the structure constant  $C_n^2$ as described in the Kolmogorov turbulence theory \cite{Tatarski1961}. This quantifies the amount of refractive index change a beam undergoes when traveling through a turbulent section, and thus how much the beam is deflected and wanders. In the atmosphere, typical values are $C_n^2 = 10^{-16} - 10^{-12}$ m$^{-2/3}$ for calm and stormy conditions respectively \cite{Bendersky2004}. On the other hand, because different parts of the beam travel through different turbulence cells, the transverse coherence length of the beam $r_0$ is affected and the beam breaks up into several coherent parts (speckles). The transverse coherence length, or Fried parameter, is the diameter over which the rms fluctuations of the phase remain below a 1 radian threshold \cite{Paunescu2009} and is defined as \cite{Fried1966} 
\begin{equation}
r_0 = (0.423k_0^2C_n^2 \Delta z)^{-3/5}
\label{eq_r0}
\end{equation}
Where $\Delta z $ is the length of the turbulent region, and $k_0$ the wave vector of the pulse.

A remarkable feature of filaments is that once formed, they are rather robust to turbulence \cite{Ackermann2006}. However, in the stage before the filament formation, the interaction with turbulence can have the seemingly opposite effects of either decreasing the filamentation probability \cite{Ackermann2006,Salame2007,Houard2008}, or increasing the number of filaments in a multiple-filaments regime ($P \gg P_{NL}$) \cite{Paunescu2009,Garnier2006}  and either increasing \cite{Ackermann2006} or decreasing the onset distance \cite{Kandidov1999,Shlenov2009}, depending on the initial experimental conditions.

Here, we show a different regime of turbulence-pulse interaction. Namely, a strongly turbulent environment can trigger filamentation for a laser beam that does not have enough power to filament in a calm atmosphere. Where the resilience of single filaments to turbulence is the intuitive concern for applications, we show that the opposite effect, the creation of filaments due to turbulence, is also possible. This occurs via the seeding of modulation instability (MI) by turbulence, i.e. the rapid growth of a transverse perturbation in the beam, causing the onset of single filaments for a beam that is below the power threshold to filament. While MI is commonly associated with multiple-filaments, we define the conditions for this turbulence-induced single-filaments regime, where the ratio of the coherence length to the beam diameter is a critical parameter.

\section{\label{sec:Experimental}Experimental Setup}
\begin{figure}
\centering
\includegraphics[width=9 cm]{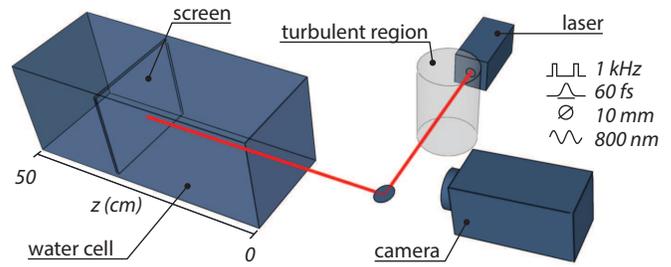}
\caption{Experimental setup. Turbulence is applied to the collimated beam before entering the water-cell. The distance traveled from the beginning of the cell to the measurement screen is denoted by $z$. The image on the measurement screen is recorded by a high speed camera capturing each individual laser shot.}
\label{Fig_Setup}
\end{figure}

As schematically depicted in Figure \ref{Fig_Setup}, our setup consists of a collimated femto-second laser, propagating through a small turbulent air region followed by a water-cell. Because the nonlinear refractive index of water \cite{Couairon2007} is about three orders of magnitude higher than of air, this table-top experiment allows us to see effects that would otherwise require long range propagation in air. 

The experiment relies on a titanium:sapphire chirped-pulse amplification laser chain producing a 60 fs pulse centered at 800 nm with a Gaussian profile of 10 mm $1/e^2$  in diameter at a repetition rate of 1 kHz. The pulse energy was reduced by lowering the amplifier pump energy. Experiments were performed at pulse energies of 0.85 mJ ($P_{\textrm{in}}$~=~14.2~GW), and 1.6 mJ ($P_{\textrm{in}}$~=~26.7~GW), as measured at the laser output by a bolometer.

The turbulent region was formed in a 12.5 cm  diameter metal cylinder, fed by a hot air blower. The strength of the turbulence was varied by changing the temperature of the hot air blower to one of three settings, corresponding to $C_n^2 = 0.7 \times 10^{-9}$ m$^{-2/3}$, $6.7 \times 10^{-9}$~m$^{-2/3}$, and $7 \times 10^{-9}$ m$^{-2/3}$, respectively. This turbulence strength was measured using the pointing stability of the laser at strongly reduced power and diameter, traveling 3.8 m in air, using the method described in \cite{Ackermann2006}. 

The collimated beam entered the turbulent region $\sim$~1~m after the laser output and traveled another 83 cm before entering the 50 cm long water-cell through a high efficiency anti-reflection 3 mm window. In our conditions the Marburger formula~\cite{Marburger1975} predicts that self-focusing requires 71~m to occur. As this distance strongly exceeds the almost 2~m propagation in air, self-focusing on this section can be neglected.

After a propagation distance $z$ (= 0 - 50 cm) in water, the beam reached a Teflon® screen where it was imaged on a single shot basis at a small angle ($\sim$ 10$\degree$) by a fast camera (Phantom v7.3) used at 1 kHz frame rate, and providing 600 x 800 pixel images. The image was restricted to the 350-600 nm spectral region by a Schott BG7 glass filter in order to filter out the near-infrared photon bath. Alternatively, images of the beam on the screen without filter were used to define the beam center position, which was subsequently used as a reference for the position of filaments within the beam profile.

The single-shot images were processed off-line. The occurrence of laser filaments at the screen location was characterized by the occurrence of a visible light on the image. Because the beam was collimated, any white light generation is due to the Kerr-self-focusing. The position of each filament relative to the beam center was recorded, in addition to its peak intensity and area. For each experimental setting 142,668 shots were recorded, consisting of 6 measurement sessions of 23,778 laser shots each. In the case of multiple filament detection, the position, peak intensity and area were recorded for each filament.

\section{\label{sec:Results}Results}
 
\begin{figure} 
\centering
\includegraphics[width=6 cm]{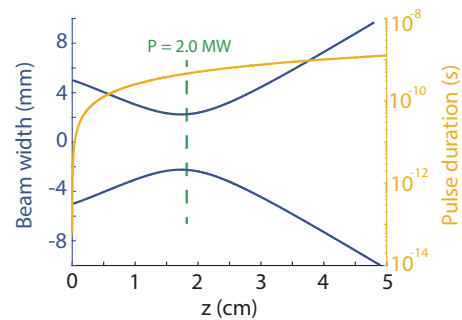}
\caption{Simulation of the pulse propagation. Because of the high dispersion the pulse length increases drastically, due to which the pulse peak power decreases at an equal rate. At the turning point of focusing to defocusing, the peak power has dropped from 14.2 GW to 2.0 MW, which is below the filamentation threshold}
\label{Fig_Simulation}
\end{figure}
 
At a pulse power of $P_{\textrm{in}}$ = 14.2 GW and in an unperturbed atmosphere, filaments are absent in the water-cell. This may seem surprising as the incident power widely exceeds the $P_{\textrm{NL,water}}$ = 6.5 MW \cite{Apeximov2015}  for $\lambda$ = 800 nm. Indeed, experiments in water are commonly performed at much lower power. However, a focusing lens is used in these cases. For our collimated beam, while $P_{\textrm{in}}$ = 2400 $P_{\textrm{NL}}$, full radial nonlinear Schr{\"o}dinger equation simulations show a divergence of the beam rather than a collapse to a filament. In water, the strong dispersion that has to be overcome on top of diffraction can cause the focusing to be arrested before collapse \cite{Fibich1996,Weiss2012}, a balance also occurring in the description of X-wave filamentation in water \cite{Faccio2006}. This interplay is captured by the model for pulse propagation in highly dispersive media by Fibich in \cite{Fibich1996}, the result of which is illustrated in Figure \ref{Fig_Simulation} for our parameters. The combination of a collimated beam, short pulse and large beam radius cause that in fact, $P_{\textrm{in}} = 14.2$  GW is below the power threshold for filamentation: $P_{\textrm{in}} < P_\textrm{{TH}}$. Note that for the same parameters, simple use of the Marburger formula \cite{Marburger1975} incorrectly predicts $L_{\textrm{SF}}$  = 2 m for the collapse-, or filament onset-length in water. Our result is in agreement with the experiments performed by Apeximov \textit{et al.} \cite{Apeximov2015}, who find that their collimated beam with similar properties produces single filaments at $P_{\textrm{in}} \sim$ 3100  $P_{\textrm{NL}}$ in water, and multiple filaments at higher powers.

Table \ref{Tab_ProbCn2} shows the filament probability $p(\geq1)$, the probability of the laser shot producing one or more filaments, for different values of $C_n^2$ at distance $z$~=~18~cm. Increasing the turbulence strength surpasses the filamentation threshold, and filaments are sporadically observed; the filament probability increases from absence to 3\% to 6\% with increasing $C_n^2$. At increased $P_{\textrm{in}}$~=~26.7~GW, the beam is predicted not to focus either in a calm atmosphere. This doubling of the pulse power generates filaments at the lowest turbulence strength $C_n^2 = 0.7 \times 10^{-9} m^{-2/3}$ and increases $p(\geq1)$ by a factor of $\sim$ 6. About 10 \% of the filamenting shots produced multiple filaments. 

\begin{table}[b]
\caption{\label{Tab_ProbCn2}%
Filamentation probability $p (\geq1)$ in \%  for different turbulence strength and beam power.
}
\begin{ruledtabular}
\begin{tabular}{lccc} 
&\multicolumn{3}{c}{$C_n^2$ ($\times$ 10$^{-9}$ m $^{-2/3})$}\\
 \cmidrule(r){2-4}
    $P_{\textrm{in}}$    & 0.1  		& 6.7  		& 7.1 \\
    \hline
    14.2 GW     & -   & 3.14 ($\pm$0.23)  	& 6.12 ($\pm$0.29)    \\
    26.7 GW     & 0.03 ($\pm$0.03)     & 16.64($\pm$0.49)& 44.28 ($\pm$1.02)
   \\
\end{tabular}
\end{ruledtabular}
\end{table}

\begin{figure}
\begin{minipage}{.5\linewidth}
\subfloat[]{\label{Fig_ProbDistCn2}\includegraphics[width=4 cm]{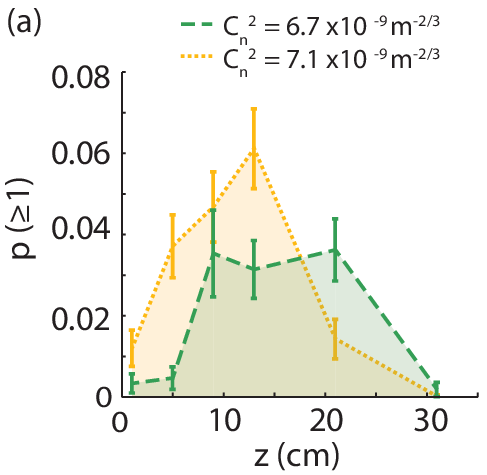}}
\end{minipage}%
\begin{minipage}{.5\linewidth}
\subfloat[]{\label{Fig_ProbDistYoung}\includegraphics[width=4 cm]{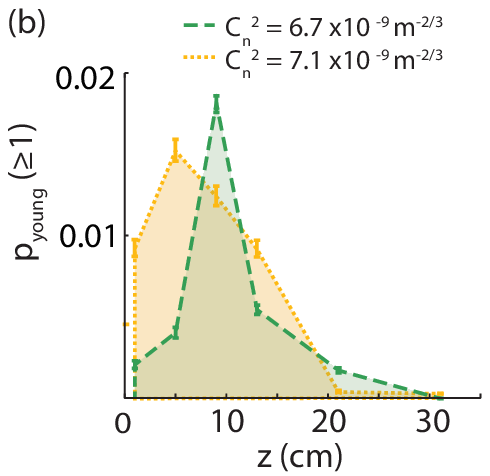}}
\end{minipage}\par\medskip
\begin{minipage}{.5\linewidth} 
\subfloat[]{\label{Fig_ProbDistAge}\includegraphics[width=4 cm]{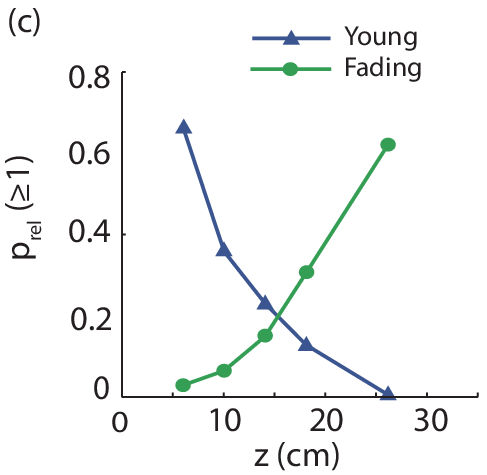}}
\end{minipage}
\caption{(a) Filament probability $p(\geq1)$ as a function of propagation distance. (b) Looking only at the young filaments $p_\textrm{young}(\geq1)$ gives an indication of the filament onset distance  (c) The relative occurrence $p_\textrm{rel}(\geq1)$ of young and dying filaments observed at each propagation distance.}
\label{Fig_ProbDistTot}
\end{figure}

\begin{figure}
\subfloat[z = 10 cm\label{Fig_example_d10_T2}]{%
  \includegraphics[width=4 cm]{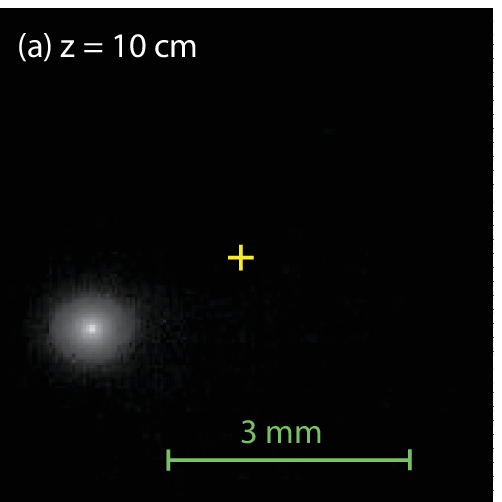}%
}\hfill
\subfloat[z = 36 cm\label{Fig_example_d36_T2}]{%
  \includegraphics[width=4 cm]{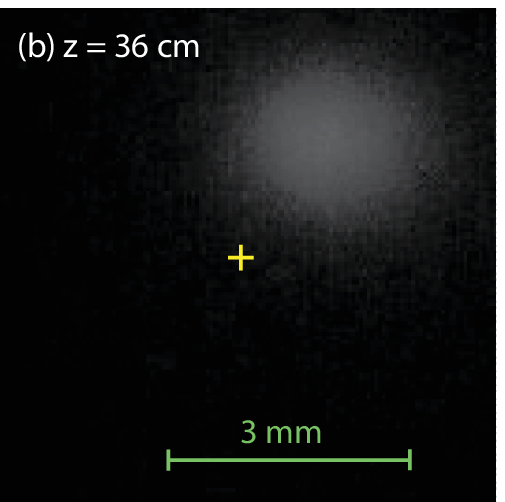}%
}
\caption{Examples of filaments as recorded by the camera for $P_{\textrm{in}}$ = 14.2 GW, $C_n^2 = 6.7 \times 10^{-9}$ m$^{-2/3}$  (a)  'Young' filament at z = 10 cm, (b) 'Fading' filament at z = 36 cm. The cross denotes the beam center.}
\label{Fig_FilPhoto}
\end{figure}

Figure \ref{Fig_ProbDistCn2} displays the filament probability as a function of distance for an incident power of 14.2 GW. Increasing the turbulence from 6.7 to 7.1 x $10^{-9}$ m$^{-2/3}$ shifts the filamenting region $\sim$ 5 cm upstream, consistent with the theoretical predictions by Pe\~nano \textit{et al.} \cite{Penano2014} that the onset distance shortens with increasing turbulence. To better characterize the filament onset distance, we analyzed the images to distinguish newly formed filaments from already fading ones. Due to the divergence of the conical emission \cite{Kosareva1997}, we expect the first ones to be characterized by a small area and high intensity (Figure \ref{Fig_example_d10_T2}, $z$ = 10 cm), and the latter by larger area and lower intensity (Figure \ref{Fig_example_d36_T2}, $z$ = 36 cm). To quantify the stage of development, the intensity criterion is used. Note that the intensity on the image does not directly correspond to the fluence of the filament, and thus it only yields a qualitative indication. Figure \ref{Fig_ProbDistAge} shows the relative occurrence of 'young' ($I>0.8I_\textrm{max}$) and 'fading' ($I<0.3I_\textrm{max}$) filaments, where $I_\textrm{max}$ is the maximal intensity recorded over all measurement sessions. Following the occurrence probability of the young filaments in Figure \ref{Fig_ProbDistYoung}, this result qualitatively agrees with the theoretical prediction made by Pe\~nano \textit{et al.} \cite{Penano2014} for the filament onset probability, where we see a localized increase and decay pattern that shifts upstream with increasing turbulence strength. 

\begin{figure}  
\centering
\includegraphics[width=4 cm]{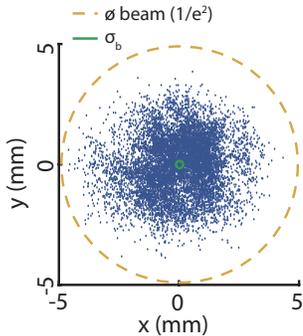}
\caption{Filament positions at $z$ = 10 cm, $P_{\textrm{in}}$ = 14.2 GW, and  $C_n^2 = 6.7 \times 10^{-9}$ m$^{-2/3}$ based on 142,668 shots for which 5,228 filaments were recorded.}
\label{Fig_Scatter}
\end{figure}

Figure \ref{Fig_Scatter} shows the transverse position on the measurement plane at distance $z = 10$ cm, of 5,288 filaments out of 142,668 laser shots. The wander of the beam center due to turbulence ($\sigma_b = 0.07$ mm) is 15 times smaller than the wander of the filaments ($\sigma_f = 1.1$ mm), indicating most filaments are nucleated off-center. The dashed line represents the beam diameter at $1/e^2$ . 

The filament position on the transverse plane as displayed by Figure \ref{Fig_Scatter}, can be described well by a Rayleigh distribution for all experimental settings, confirming the simulations performed by Chin \textit{et al.} \cite{Chin2002}. In contrast, Houard \textit{et al.} report a better fit with the Weibull distribution \cite{Houard2008}. As our system involves the study of solitons, which can be considered building blocks of rogue waves, this distinction between the distributions is relevant in the discussion of rogue wave statistics. Rogue waves are associated with long-tailed distributions. Therefore, they do not correspond to a Rayleigh distribution \cite{Dudley2014}, which stems from the angular integration of a two-dimensional normal distribution in polar coordinates. 

For the data, the normalized probability density function is defined by $PDF(r_m) = N (r_m) / N_{tot}$, where $m$ is the bin number and $N$ the filament count. The cumulative distribution $\Psi (r_m) = \sum_{i=1}^{m} PDF(r_i) $. Conversely, $PDF(r) = d\Psi/dr$. Both the Rayleigh and Weibull distribution can be parameterized as follows

\begin{equation}
\Psi(r) = 1-e^{-(r/w)^b}
\label{Eq_Psi}
\end{equation}

\begin{figure}
\subfloat[\label{Fig_PDF}]{%
  \includegraphics[width=4 cm]{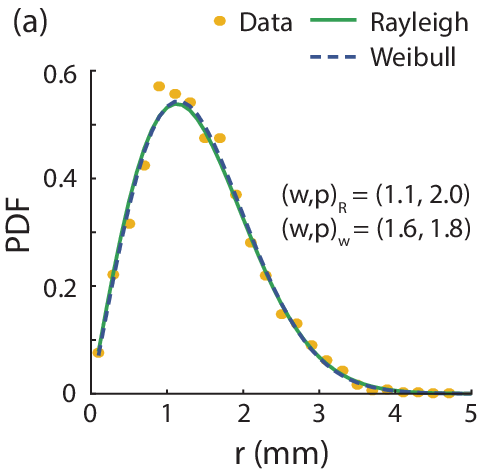}%
}\hfill
\subfloat[\label{Fig_SSE}]{%
  \includegraphics[width=4 cm]{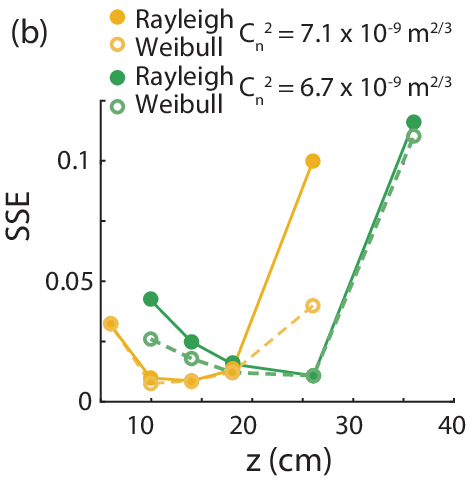}%
}
\caption{(a) Example of the probability density function for $ z = 10 $ cm at $C_n^2 = 6.7 \times 10^{-9} m^{-2/3}$. (b) Sum of Squared errors. Experimental settings that contained too few data points were excluded from the analysis.}
\label{Fig_PDFTot}
\end{figure}

For the Rayleigh distribution, $w$ characterizes the width of the distribution and $b \equiv$ 2. In the 2-parameter Weibull distribution, both $b$ and $w$ are free parameters. Obviously, as the Weibull distribution is a generalization of the Rayleigh one, it can only give an equal or better fit.  
In our case, the Weibull fit is only marginally better. More specifically, Figure \ref{Fig_PDFTot} shows that while the Weibull PDF consistently gives a slightly better fit, both distributions have very similar residuals (Figure~\ref{Fig_SSE}).  Indeed, the fits are almost superimposed (Figure~\ref{Fig_PDF}), showing that the deviation from a normal distribution is marginal.

\section{\label{sec:Discussion}Discussion}

We observe turbulence-induced single filaments, and a decrease of the filament onset distance with increasing turbulence. To some extent, this can be placed in the theoretical framework detailed by Pe\~nano \textit{et al.} \cite{Penano2014}, who account for the seemingly contradicting results mentioned in the introduction of turbulence either lengthening or shortening the filament onset distance $L'_{\textrm{SF}}$ and either increasing or decreasing filament probability. More specifically, they distinguish three regimes of turbulence-beam interactions based on the ratio of coherent area in the beam and the power it contains, over the nonlinear power to overcome in order to self-focus:  

\begin{equation}
Q^2(z) = \frac{6 P r_0^2 (z)}{P_{NL} R_0^2}
\label{Eq_Q}
\end{equation}

Where  $r_0$ is the transverse coherence length, given by Eq. (\ref{eq_r0}) and $R_0$ the beam radius. For $Q^2 \gg  1 $ (i), the turbulence is weak. A Gaussian-profiled beam collapses on-axis and produces a single filament. The turbulence acts as noise that decreases the probability of whole beam self focusing (WBSF) and lengthens the self-focusing distance $L'_{\textrm{SF}}$ . For $Q \geq 1$ (ii) the turbulence increases while $P \gg  P_{\textrm{NL}}$, and splits up the beam into smaller coherent areas of the order of $P_{\textrm{NL}}$. Because the nonlinear system is sensitive to transverse perturbations, modulation instability (MI) amplifies cells having a power $\sim P_{\textrm{NL}}$, hereby nucleating filaments. The maximal MI amplitude occurs when $Q \sim 1 $. The idea that turbulence can seed MI \cite{Shlenov2009,Paunescu2009,Garnier2006}, has been experimentally demonstrated by Paunescu \textit{et al.} \cite{Paunescu2009}, who find a shortening of the focal distance, and an increase in the number of filaments due to turbulence in the multi-filament regime. Lastly, if $Q < 1$ (iii) turbulence is so strong that the beam is incoherent over any area containing $P_{\textrm{NL}}$ so that turbulence suppresses both MI and WBSF, and thus filament formation.

Because of the strong dispersion effect on a collimated beam in water, we interpret Eq.
(\ref{Eq_Q})  based on $P_{\textrm{TH}}$ rather than $P_{\textrm{NL}}$, in which case we have $P_{\textrm{in}}$ = (1-$\delta$)$P_{\textrm{TH}}$ , where $\delta$ is small, as we are slightly under the filamenting threshold. For our setup $r_0 \approx 2.5$~mm, and thus $r_0 \sim R_0/2$. Consequently, we obtain $Q \sim 1 $, and our results fall in case (ii): the MI driven regime. This is supported by the fact that we observe a shortening of the focus distance, an increase in filamentation probability, and off-axis filaments due to turbulence. 

Our results present two novel findings not featured in previous work. Firstly, our results show that the MI interpretation even holds in the single filament regime, which is not commonly associated with MI. The previously mentioned studies \cite{Penano2014,Shlenov2009,Paunescu2009,Garnier2006}, are performed with $P \gg P_{\textrm{NL}}$, corresponding to the multi-filament regime. Indeed, since one condition in Pe\~nano's model is $r_0 \ll R_0$, when strictly applied, formula \ref{Eq_Q} dictates that $P/P_{\textrm{NL}}$ must be quite high to obtain $Q \sim 1$.  Note that this is only a qualitative estimation. Here, in contrast, we observe the MI driven regime while we are not strictly in the range $r_0 \ll R_0$. By this experimental demonstration we extend the domain of validity, and show it is possible to be in the single filament regime and observe MI induced filaments. Secondly, the effect we observe is beyond a mere shifting of the filament onset distance $L'_{\textrm{SF}}$. Rather, it demonstrates a qualitative change from absence to presence of filaments. 

Our findings are not specific to our configuration, where the turbulent medium was air and the non-linear propagation occurred in water. While the use of air as the turbulent medium is convenient from a practical point of view, it has no influence on our results, since, as detailed above, the propagation of our beam is mainly linear over the 2~m of propagation in this medium.

\section{Conclusion}
In conclusion, we have observed the nucleation of single filaments induced by turbulence in an otherwise non-filamenting beam. We hereby demonstrate that a turbulent region is not only capable of reducing the filament onset distance or increasing the number of filaments, but can act as a switch between presence and absence of a single filament. This implies that the refractive index change caused by the turbulent region is of such magnitude that it can induce modulation instability in a beam that would not have enough power to filament. We theorize that the size scale of the coherence length being of the same order of magnitude as the beam diameter is the key point in this observation. The turbulence-generated filament position on the transverse plane follows a Rayleigh distribution.


\section*{Acknowledgments}
We acknowledge financial support
from the Swiss National Science Foundation (project 200021-155970). N.B. and J.P.W. acknowledge financial support from the European Research Council advanced grant « Filatmo », S.H. from the Cofund program of the FP7, and S.H. and J.G. from the NCCR (National Center of Competence in Research) “MUST” (Molecular Ultrafast Science and Technology) from the Swiss NSF.  The authors thank Michel Moret for his technical help.

\bibliography{TurbulenceBib}

\end{document}